\documentclass[10pt]{iopart}
\usepackage{epsfig}
\begin{document}

\title[off resonance absorption and dispersion]{Off-resonance absorption and dispersion in a Doppler-broadened medium}
\author{Paul Siddons, Charles S Adams and Ifan G Hughes}

\address{Department of Physics,
Durham University, South Road, Durham, DH1~3LE, UK}
\ead{paul.siddons@durham.ac.uk}
\date{\today}
\begin{abstract}
\noindent We study the absorptive and dispersive properties of Doppler-broadened atomic media as a function of detuning.  Beginning from the exact lineshape calculated for a two-level atom, a series of approximations to the electric susceptibility are made.  These simplified functions facilitate direct comparison between absorption and dispersion, and show that dispersion dominates the atom-light interaction far from  resonance.  The calculated absorption and dispersion are compared to experimental data, showing the validity of the approximations. 
\end{abstract}

\section{Introduction}\label{sec:intro}

In his paper we investigate the relationship between the absorption and dispersion experienced by off-resonant radiation interacting with an inhomogeneously broadened atomic medium.  Of particular interest is the response of the medium when the detuning is larger than the inhomogeneous linewidth.  In this region scattering of photons is reduced, but this does not necessarily mean that the dispersive atom-light coupling suffers accordingly.  The large dispersion of off-resonant Doppler-broadened systems has been exploited in a number of studies.  Slow-light experiments~\cite{Vanner08,Camacho07} and theoretical studies~\cite{Shakh08} utilise the large group refractive index associated with the medium to control the propagation of broadband optical pulses. Slow-light
interferometers  using  monochromatic~\cite{ShiOptLett, Purves} and broadband light~\cite{Shi07} have also been demonstrated.  The off-resonant Faraday effect can be used to separate the sidebands of Raman light with high fidelity~\cite{Abel09}, and can be used as a dispersive probe with continuous-wave or pulsed light~\cite{Siddons09}. The non-invasive nature of off-resonant dispersive probing could also lead to the possibility of ``weak'' measurements~\cite{Ahanorov88} in the context of quantum non-demolition (QND).  In many of these experiments there is a trade-off between the absorption of the optically-thick medium and the magnitude of the effect of interest.  As the real and imaginary parts of the medium's susceptibility show different spectral dependences it is not obvious 
at which detuning to perform the experiments. The motivation of this study is to take previously developed analytic results for the susceptibility~\cite{Siddons08} and to investigate the domain of validity of two approximations which facilitate the analysis of experimental data.

The structure of this paper is as follows.  In section~\ref{sec:chi} we describe the function which governs the absorption and dispersion of a Doppler-broadened medium, and go on to make approximations to this function.  We then use these analytic approximations to compare the absorptive and dispersive characteristics of an atomic resonance.  In section~\ref{sec:results} we compare the theoretical expressions to experiment, and in section~\ref{sec:conclusion} we draw our conclusions.

\section{The electric susceptibility}\label{sec:chi}

The electric susceptibility of a medium, $\chi$, describes the medium's absorptive and dispersive properties.  For the case of an isolated resonance in a Doppler-broadened atomic medium the susceptibility as a function of detuning from resonance, $\Delta$, is~\cite{Siddons08}: 
\begin{eqnarray}
\chi(\Delta) &=& c_{\mathrm{m_F}}^2\frac{d^2\mathcal{N}}{\hbar \epsilon_0}s(\Delta).\label{eq:chi}
\end{eqnarray}
Here $c_{\mathrm{m_F}}^2$ is the transition strength factor for the transition \mbox{$\left|F_{\mathrm{g}},m_{\mathrm{Fg}}\right\rangle\rightarrow\left|F_{\mathrm{e}},m_{\mathrm{Fe}}\right\rangle$}, \mbox{$d=\langle L_{\mathrm{g}}||e\textrm{\textbf{r}}||L_{\mathrm{e}}\rangle$} is the reduced dipole matrix element of the \mbox{$|L_{\mathrm{g}}\rangle\rightarrow |L_{\mathrm{e}}\rangle$} transition, $\mathcal{N}$ is the atomic number density of state $\left|F_{\mathrm{g}},m_{\mathrm{Fg}}\right\rangle$, and $s(\Delta)$ is the lineshape factor.  The total susceptibility of the medium is obtained by summing over all transitions which the light is stimulating.  The absorption coefficient is proportional to the imaginary part of the susceptibility, $\chi^{\mathrm{I}}$, and has the form of the well-known Voigt profile.  Dispersion results from the real part, $\chi^{\mathrm{R}}$.

The lineshape factor $s(\Delta)$ is the convolution of $f(\Delta)$, the homogeneous atomic lineshape, and $g(v)$, the Gaussian distribution of velocities $v$.  This convolution is given by:
\begin{equation}
s(\Delta) =\int ^{+\infty}_{-\infty} f(\Delta-kv)\times g(v) dv,
\label{eq:vintegral}
\end{equation}  
where $k$ is the wavenumber of the radiation, and
\begin{eqnarray}
f(\Delta)&=&\frac{i}{\Gamma/2-i\Delta},\label{eq:convl}\\
g(v)&=&\frac{1}{u\sqrt{\pi}}\mathrm{e}^{-(v/u)^2}.\label{eq:convg}
\end{eqnarray}
Here $\Gamma$ is the FWHM (Full-width at half-maximum) of the homogeneous broadening, and  $u$ is the $1/\mathrm{e}$ of the inhomogenous broadening mechanism (and the RMS atomic speed).  The lineshape $s(\Delta)$ is related to the Faddeeva (or complex error) function, $w(iz)$, of complex argument $z$, via
\begin{eqnarray}
s(\Delta) &=& \frac{i\sqrt{\pi}}{ku}w(iz),\label{eq:s}\\
w(iz)&=&\frac{i}{\pi}\int^{+\infty}_{-\infty}\frac{\mathrm{e}^{-x^2}}{iz-x}\textrm{d}x=\mathrm{e}^{z^2}\mathrm{erfc}(z),\label{eq:wiz}\\
z(\Delta)&=&\frac{1}{2}\frac{\Gamma}{ku}-i\frac{\Delta}{ku}.\label{eq:z}
\end{eqnarray}
Equation (\ref{eq:s}) is the exact analytic lineshape of the Doppler-broadened susceptibility; unfortunately this exact result can be difficult to use.  Although algorithms exist for the Faddeeva and complementary error function $\mathrm{erfc}(z)$, they are not easy to manipulate analytically, and can be time-consuming to evaluate numerically.  Consequently it is difficult to relate $\chi$ to $z$ and the parameters of which they are composed (namely the widths $\Gamma$ and $ku$).  This in turn makes it difficult to see the relationship between the absorptive and dispersive properties.  The preceding reasons motivate our study of approximations to the analytic result by looking at the Faddeeva function in two regimes, named the Gaussian and Lorentzian approximations for reasons that will become apparent. 

\subsection{The Gaussian approximation}\label{sec:gapprox}

We consider the situation where the broadening due to atomic motion is much larger than natural broadening, which is the case for typical room temperature alkali-metal atoms.  For this approximation we therefore look at the limit that $\Gamma/ku\rightarrow0$ in the derivation of the Faddeeva function.  Starting from the homogeneous lineshape of equation~(\ref{eq:convl}),
\begin{equation}
\lim_{\Gamma/ku\rightarrow0}f(z)=-\frac{ku}{\Delta}+i\pi\delta(\frac{\Delta}{ku}),
\end{equation}
where the imaginary part is given by a Dirac delta function.  With this expression substituted into equation~(\ref{eq:s}), the real and imaginary parts of the susceptibility are, respectively
\begin{eqnarray}
s^{\mathrm{R}}&=&-\frac{\sqrt{\pi}}{ku}\:\mathrm{e}^{-(\Delta/ku)^2}
\mathrm{erfi}(\Delta/ku),\label{eq:gsr}\\
s^{\mathrm{I}}&=&\ \ \ \: \frac{\sqrt{\pi}}{ku}\:\mathrm{e}^{-(\Delta/ku)^2}.
\label{eq:gsi}
\end{eqnarray}
The real part contains the imaginary error function $\mathrm{erfi}(z)$ which is similar to the Faddeeva function in that it needs to be evaluated numerically.  The imaginary term is the convolution of a Gaussian and a Dirac delta function, and as expected evaluates to the Gaussian function responsible for Doppler-broadening, with the FWHM Doppler width $\Delta\omega_{\mathrm{D}}=2\sqrt{\ln2}\:ku$.

In this approximation both $s^{\mathrm{R}}$ and $s^{\mathrm{I}}$ have a Gaussian detuning dependence, whose exponential decrease means that it decays rapidly away from resonance.  $s^{\mathrm{R}}$ has an additional imaginary error function dependence which increases rapidly with detuning; hence dispersion contains the long-range characteristics associated with the Faddeeva function.  Thus the imaginary part will only be valid close to resonance, whereas the real part of the Gaussian approximation is expected to be in good agreement with the Faddeeva function for all detunings

\subsection{The Lorentzian approximation}\label{sec:lapprox}

In the Gaussian approximation we made the assumption that homogeneous broadening was negligible compared to inhomogeneous broadening, based on the ratio of their frequency widths.  We saw that this is not true far from resonance.  In this section we will find regimes under which homogeneous dominates the susceptibility.  We begin by noting that the complementary error function can be written in the form of a continued fraction~\cite{ContFrac1,ContFrac2}
\begin{equation}
\sqrt{\pi}\:\mathrm{erfc}(z)= 2\int ^{\infty}_{z}\mathrm{e}^{-t^2}\mathrm{d}t=\frac{2\mathrm{e}^{-z^2}}{2z+\frac{2}{2z+\frac{4}{2z+\frac{6}{2z+\ldots}}}}.
\label{eq:efn}
\end{equation}
For $|z|\gg1$ the continued fraction can be approximated to $\mathrm{e}^{-z^2}/z$. This requires either of the following conditions to be fulfilled:

\begin{tabular}{cc}
(\textit{i})& $|\Delta|\gg\Delta\omega_{\mathrm{D}}$\\
(\textit{ii})& $\;\Gamma\ \gg\Delta\omega_{\mathrm{D}}$
\end{tabular} 

\noindent The first condition is that the laser is detuned from resonance further than the Doppler width and is essentially a property of the light source; the second is that natural broadening dominates over Doppler broadening and is a property of the medium.  The Doppler width can be reduced by, for example, using cold atoms at sub-milliKelvin temperatures.  Many experiments, including the ones considered in this paper, are conducted with alkali-metal atoms on the D-line at room temperature (or hotter); the parameters of interest are then $\Gamma\sim 2\pi\times 5$~MHz, and $\Delta\omega_{\mathrm{D}}\sim 2\pi\times 0.5$~GHz, thus $\Gamma/\Delta\omega_{\mathrm{D}}\sim10^{-2}$.  Therefore, for the limit $|z|\gg1$ to be valid, it is necessary to be detuned far from resonance, $|\Delta|\gg\Delta\omega_{\mathrm{D}}$. 

Substituting the approximated $\mathrm{erfc}(z)$ into (\ref{eq:s}) we get the result
\begin{eqnarray}
s&=&\frac{i}{ku}\frac{1}{z}=\frac{i}{\Gamma/2-i\Delta}.
\label{eq:lapprox}
\end{eqnarray}
Note that this is identical to the case for homogeneous broadening, e.g. an ensemble of stationary atoms, or atoms at ultralow temperatures for which Doppler broadening is negligible~\cite{Shultz2008}. The real part of the susceptibility gives the dispersion function, and the imaginary part the Lorentzian function; specifically
\begin{eqnarray}
s^{\mathrm{R}}&=&-\frac{\Delta}{(\Gamma/2)^2+\Delta^2},\label{eq:lsr}\\
s^{\mathrm{I}}&=&\frac{\Gamma/2}{(\Gamma/2)^2+\Delta^2}.
\label{eq:lsi}
\end{eqnarray}
Furthermore, since $|\Delta|\gg\Delta\omega_{\mathrm{D}}\gg\Gamma$, these relations simplify further to $s^{\mathrm{R}}=-1/\Delta$, and $s^{\mathrm{I}}=\Gamma/2\Delta^{2}$ respectively.  These detuning dependences are discussed further in section~\ref{sec:comabsdis}.

The physical interpretation of the Lorentzian approximation is that the Gaussian lineshape responsible for inhomogeneous broadening decreases exponentially with detuning, whereas the homogeneous lineshape decreases much more slowly in the wings.  Hence the contribution to the overall lineshape far from resonance is dominated by the Lorentzian function, and both absorption and dispersion will be well approximated.

\subsection{Validity of the approximations}\label{sec:val}

\begin{figure}[p!]
\centering
\includegraphics[width=8.5cm]{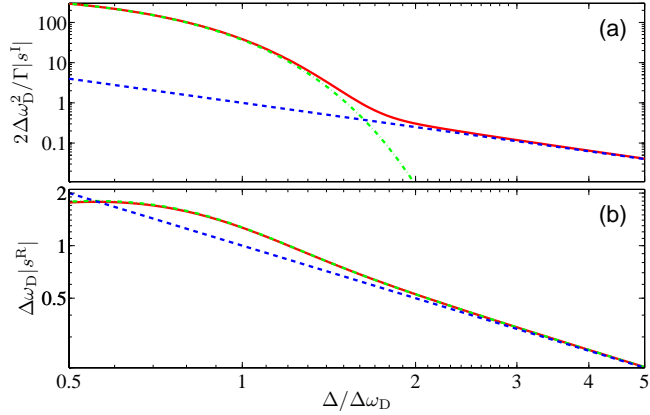}
\caption{Comparision of the Faddeeva function and its approximations.  The real and imaginary parts of the susceptibility lineshape, $s$, are shown in parts (a) and (b), respectively.  The horizontal axis is detuning, $\Delta$, in terms of the Doppler width, $\Delta\omega_{\mathrm{D}}$.  The solid red curve shows $s$ calculated using the Faddeeva function (with $\Gamma/\Delta\omega_{\mathrm{D}}=10^{-2}$), whilst the Gaussian and Lorentzian approximations are shown as dot-dashed green and dashed blue curves, respectively.}
\label{fig:sapprox}
\end{figure}

Figure~\ref{fig:sapprox} shows the lineshape, $s$, of the Faddeeva function and its Gaussian and Lorentzian approximations, for a typical room temperature alkali-metal atomic ensemble where the Doppler-broadening is two orders of magnitude larger than natural broadening.  It can be seen in figure~\ref{fig:sapprox}(a) that for $|\Delta|<1.5\times\Delta\omega_{\mathrm{D}}$ the imaginary part of the Faddeeva function is adequately described by the Gaussian approximation, and for $|\Delta|>2\Delta\omega_{\mathrm{D}}$ the Lorentzian approximation holds.  Therefore, close to resonance, Doppler-broadening dominates the absorptive interaction; whereas natural broadening dominates at large detuning.  A similar situation for the real part of the Lorentzian aproximation is seen in figure~\ref{fig:sapprox}(b), i.e. it is valid for $|\Delta|>2\Delta\omega_{\mathrm{D}}$.  However, the Gaussian approximation is in good agreement with the Faddeeva function over the whole spectral range, as predicted in section~\ref{sec:gapprox}.    

\subsection{Comparing absorption and dispersion}\label{sec:comabsdis}

\begin{figure}[p!]
\centering
\includegraphics[width=8.5cm]{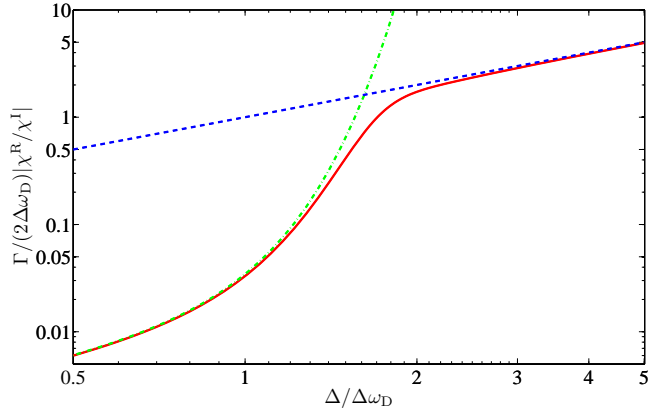}
\caption{The relative importance of dispersive and absorptive properties of a Doppler-broadened atomic medium.  The ratio $|\chi^{\mathrm{R}}/\chi^{\mathrm{I}}|$ for a single transition is shown as a function of detuning, $\Delta$, in terms of the Doppler width, $\Delta\omega_{\mathrm{D}}$.  The solid red curve shows $\chi$ calculated using the Faddeeva function (with $\Gamma/\Delta\omega_{\mathrm{D}}=10^{-2}$), whilst the Gaussian and Lorentzian approximations to $\chi$ are shown as dot-dashed green and dashed blue curves, respectively.}
\label{fig:chiapprox}
\end{figure}

Figure~\ref{fig:chiapprox} shows the ratio $|\chi^{\mathrm{R}}/\chi^{\mathrm{I}}|$, calculated using the Faddeeva function and its Gaussian and Lorentzian approximations.  It shows that the ratio between dispersion and absorption continually increases with detuning, with the asymptotic limit $2|\Delta|/\Gamma$ from the Lorentzian approximation.  Note, however, that dispersion also decreases linearly in this limit.  Hence, any dispersive effects which require low absorption are best performed far from resonance under conditions which increase the atom-light interaction e.g. high atomic density~\cite{Siddons09} or stronger coupling~\cite{Kubasik09}.  

\subsection{Hyperfine structure}

We have shown that Doppler-broadening can effectively be ignored for detunings $|\Delta|>2\Delta\omega_{\mathrm{D}}$.  However, this situation is somewhat complicated due to the presence of hyperfine structure.  For alkali-metal atoms the hyperfine structure is such that the ground state splitting, $\Delta\omega_{\mathrm{hfs}}$, is much larger than the room temperature Doppler width.  This is not  the case for the excited states, which tend to have intervals of comparable size to $\Delta\omega_{\mathrm{D}}$.  Hence, in order to calculate $\chi$ near to the line-centre, each individual hyperfine transition needs to be modelled individually, although for some purposes excited state splitting can be ignored (for example, on the $\mathrm{D}_2$ lines of Rb~\cite{Shi07} and Cs~\cite{Camacho07}).  Far from line-centre, at detunings larger than the ground state hyperfine splitting, it is possible to approximate all hyperfine transitions to a single Lorentzian function.  By performing this calculation we find that there is a less than 5\% error for $|\Delta|>3.5\times\Delta\omega_{\mathrm{hfs}}$. 

\section{Comparison between theory and experiment}\label{sec:results}

In order to test experimentally  the validity of the approximations to the Faddeeva function, the transmission of a probe beam on the $\mathrm{D}_1$ line of rubidium was recorded.  The output from an external cavity diode laser at 795~nm was attenuated to be less than $1\:\mu$W such that it is in the weak probe limit (see ref.~\cite{weak}) and passed through a 75~mm heated vapour cell, based on the design of~\cite{Danny}.  A solenoid provided the heating and magnetic field, when required.  The cell contained Rb isotopes according to the ratio  $^{87}$Rb:$^{85}$Rb of 99:1. 

\subsection{Absorption}\label{sec:abs}

\begin{figure}[p!]
\centering
\includegraphics[width=8.5cm]{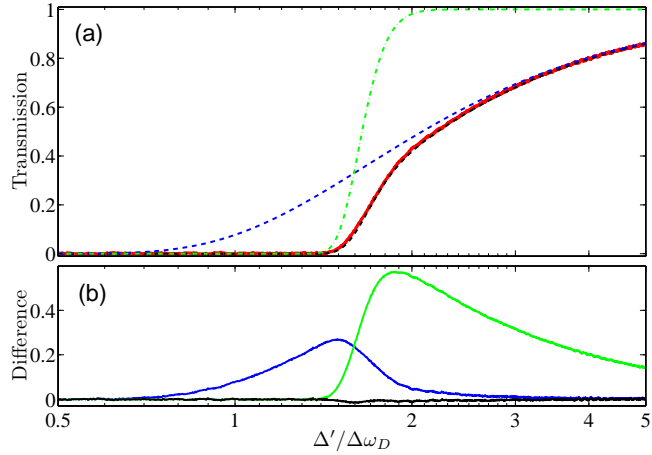}
\caption{Comparison between experiment and theory for the transmission of a weak probe beam through a vapour cell (a)  Experimental data are shown in red,  whilst the dashed black curve shows the transmission calculated using the Voigt function.  The Gaussian and Lorentzian approximations to the Voigt function are shown as dot-dashed green and dashed blue curves, respectively.  (b) The difference in transmission between theoretical and measured data.  The experimental data were obtained with red-detuned light, but plotted against  $\Delta'=-\Delta$.  The origin of the detuning axis is from the $^{87}$Rb $F_{\mathrm{g}}=2\rightarrow F_{\mathrm{e}}=1$ transition.}
\label{fig:trans}
\end{figure}

The solid curve in figure~\ref{fig:trans}(a) shows the transmission measured at $132^\circ$C as a function of detuning from the weighted line-centre in units of Doppler width, $\Delta\omega_{\mathrm{D}}=2\pi\times584\:\mathrm{MHz}$.  Absorption in the region shown is due to the $^{87}$Rb $F_{\mathrm{g}}=2\rightarrow F_{\mathrm{e}}=1,2$ transitions.  Theoretical transmission (dashed curves) was calculated using equations~(\ref{eq:s}), (\ref{eq:gsi}) and (\ref{eq:lsi}) to model each hyperfine transition involved.  The transmission difference between theory and experiment is shown in figure~\ref{fig:trans}(b).  It can be seen that the Faddeeva function agrees with the measured data to within the noise level; any discrepancy beyond this is due to the fitting of the frequency axis.  Both the Gaussian and Lorentzian approximations agree on resonance.  This is because the transition is optically thick so any variation between the two approximations is obscured.  At a detuning of about one Doppler width the Lorentzian is no longer at zero transmission and only matches with measured data again for detunings greater than two Doppler widths.  Conversely, the Gaussian approximation matches the experiment up to about 1.5 Doppler widths before differing significantly for larger detunings.  It was stated in the derivation of the Gaussian and Lorentzian approximations that Doppler broadening dominates close to resonance, whilst natural broadening dominates far from resonance.  It can be seen that at around two Doppler widths absorption due to the Gaussian lineshape rapidly decreases, and it is from this point that the Lorentzian function becomes the dominant broadening mechanism.

The situation with the two approximations is similar to that seen in previous experiments, where Gaussian fits to data are used when the line-centre is of interest, e.g. ref.~\cite{Gauss1}, and Lorentzian fits for off resonant behaviour, e.g. ref.~\cite{Camacho07}.  However, we have derived these lineshapes \textit{ab initio} and quantified their regime of validity.

\subsection{Dispersion}\label{sec:dis}

The dispersive properties of the medium are easily probed via the Faraday effect~\cite{Siddons09}.  A  rotation of the plane of polarisation arises due to the difference in the refractive indices of left and right circularly polarised light.  In a heated vapour this rotation can be as large as tens of radians over a frequency range of many Doppler widths.  Using the  technique described in ref.~\cite{Siddons09} we measured the rotation of light polarisation using a differencing signal in a balanced polarimeter.  Light transmitted through the vapour is sent through a polarisation beam splitter and the resulting vertical and horizontal polarisations directed to a differencing photodiode, where the intensities $I_{x}$ and $I_{y}$ are subtracted.  An important feature of this signal is that the period of oscillation is dependent on dispersion, whilst being independent of absorption.  From the zero crossings one can extract the rotation angle, $\theta$.  

\begin{figure}[p!]
\centering
\includegraphics[width=8.5cm]{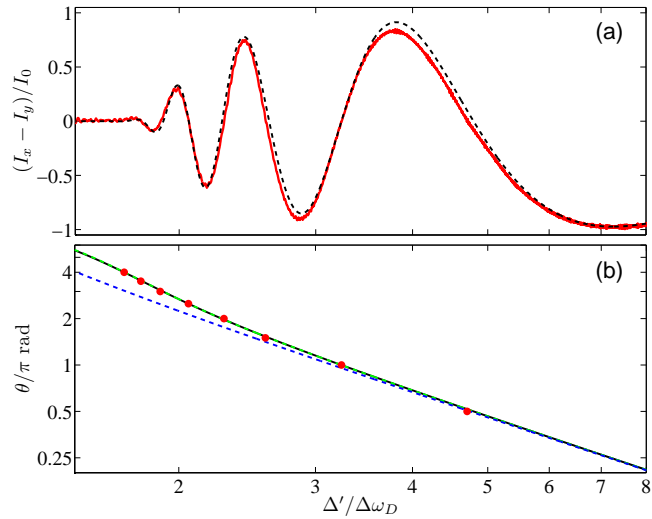}
\caption{Comparison between experiment and theory for the polarisation rotation  of a weak probe beam through a vapour cell.  (a) Balanced differencing signal.  Experimental data are shown in red,  whilst the dashed black curve shows the theoretical signal calculated using the Faddeeva function.  (b) The rotation angle of the beam's plane of polarisation: red points are from the zero crossings of the measured data, curves are calculated using the Faddeeva (black), Gaussian (green) and Lorentzian (blue) functions. The Doppler width is $\Delta\omega_{\mathrm{D}}=2\pi\times 569\:\mathrm{MHz}$.}
\label{fig:nemo}
\end{figure}

Figure~\ref{fig:nemo}(a) shows the differencing signal for an atomic temperature of $112^\circ$C and applied field of 200~G.  The signal is  normalised to the maximum intensity, $I_0$, received by one of the photodiodes in the absence of a magnetic field.  Also shown is the theoretical signal calculated by solving the complete Hamiltonian of the system, with a Faddeeva lineshape.  There is good agreement between the two curves, the main difference being in the amplitude, which is due in part to the differencing photodiodes not being perfectly balanced.  Figure~\ref{fig:nemo}(b) compares the Faddeeva function and its approximations to the rotation angle experienced by the linearly polarised probe beam. The data points are taken from the zero crossings in figure~\ref{fig:nemo}(a).  Excellent agreement is seen between measured data and both the Faddeeva function and Gaussian approximation over the whole spectral range, whilst the Lorentzian approximation only holds far from resonance, in agreement with the conclusion of sections~\ref{sec:gapprox} and~\ref{sec:lapprox}.

\section{Conclusion}\label{sec:conclusion}

We have seen in the previous two sections that the Faddeeva function gives a good fit to data over the whole frequency range, whereas the Lorentzian approximation is only valid far from resonance.  The Gaussian approximation, however, shows contrasting behaviour in its absorptive and dispersive properties.  With regard to absorption it is valid close to resonance only, whilst the dispersive properties are accounted for over all detuning.  The differing nature of the two approximations stems from the fact that only in the Lorentzian approximation did we assume that $|\Delta|\gg\Delta\omega_{\mathrm{D}}$.    

The approximations to the electrical susceptibility we have described facilitate (i) the analytic manipulation of $\chi$, allowing a comparison between the dispersive and absorptive properties of atomic media; and (ii) ease of computation of properties of interest.  In particular, we have seen that for a detuning greater than two Doppler-broadened linewidths the fully analytic Lorentzian approximation is valid.  From this we have shown that off resonance dispersion increasingly dominates over absorption. 

\ack This work is supported by EPSRC.  We thank A S Arnold for valuable discussion.

\end{document}